\documentclass[twocolumn,showpacs,preprintnumbers,amsmath,amssymb]{revtex4}
\usepackage{amssymb}
\usepackage{graphicx}
\usepackage{dcolumn}
\usepackage{color}
\usepackage{bm}
\newcommand{\be}{\begin{equation}}
\newcommand{\bea}{\begin{eqnarray}}
\newcommand{\bdm}{\begin{displaymath}}
\newcommand{\edm}{\end{displaymath}}
\newcommand{\eea}{\end{eqnarray}}
\newcommand{\ee}{\end{equation}}
\newcommand{\la}{\langle}
\newcommand{\ra}{\rangle}

\begin{document}
\title{Ground-state long-range order in quasi-one-dimensional Heisenberg
quantum antiferromagnets: High-order coupled-cluster calculations}
\author{Ronald Zinke$^1$ }
\author{J\"org Schulenburg$^2$}
\author{Johannes Richter$^1$}
\affiliation{$^1$Institut f\"ur Theoretische Physik, Universit\"at Magdeburg,
P.O. Box 4120, D-39016 Magdeburg\\$^2$Universit\"atsrechenzentrum, Universit\"at Magdeburg, P.O. Box
4120, D-39016 Magdeburg}

%
\today
%
\begin{abstract} {We investigate  the ground-state  magnetic long-range order of 
quasi-one-dimensional quantum Heisenberg antiferromagnets for spin 
quantum numbers $s=1/2$ and $s=1$. We use the coupled 
cluster method to calculate the sublattice magnetization and 
its dependence on the inter-chain coupling $J_{\perp}$. 
We find that for the unfrustrated spin-$1/2$ system, 
an infinitesimal inter-chain coupling is sufficient to stabilize 
magnetic long-range order, in  agreement with results obtained 
by other methods.
For  $s=1$, we find that a finite inter-chain coupling 
is necessary to stabilize magnetic long-range order.
Furthermore, we consider a quasi one-dimensional spin-$1/2$ system, 
where a frustrating next-nearest neighbor in-chain coupling is included. 
We find that  for stronger frustration as well, a finite inter-chain coupling 
is necessary to 
have magnetic long-range order in the  ground state, and that the strength 
of the inter-chain coupling necessary to establish 
magnetic long-range order  is
related to the size of the spin gap of the isolated chain. 
}
\end{abstract}

\maketitle

\section{\label{sec:level1}Introduction}
Low-dimensional quantum antiferromagnets have attrac\-ted much attention as model systems where strong quantum fluctuations may destroy magnetic long-range
order in the ground state (GS).\cite{book}  In particular, the one-dimen\-sio\-nal (1d) quantum Heisenberg antiferromagnet (HAFM) does not exhibit magnetic long-range order (LRO). In addition, there is a basic difference between
half-integer and integer antiferromagnetic Heisenberg chains.\cite{haldane}
While the 1d HAFM with half-integer spin quantum number exhibits a gapless 
excitation spectrum and a power-law decay of spin-spin correlations, a faster exponential decay of spin-spin correlations - accompanied by a finite excitation 
gap $\Delta$ (spin gap) - is observed for integer-spin 1d HAFM. However, it is known that for the 1d spin-half HAFM a frustrated next-nearest neighbor exchange coupling may also open an excitation gap (for a more
detailed discussion of 1d spin systems,  see  \cite{mikeska}).
For the formation of magnetic LRO in HAFM, the transition to two-dimensional (2d) lattices is crucial. The HAFM on 
2d bipartite lattices     
exhibits magnetic LRO at zero temperature and only competing interactions may destroy LRO
(for a more detailed discussion of 2d spin systems,  see \cite{manousakis,richter04}).
In real materials we are often faced with the situation that the nearest-neighbor
in-chain coupling $J_1$ is dominant but an inter-chain coupling $J_{\perp}$ 
is also present. A very weak inter-chain coupling even seems to be a rare exception, see e.g.
Ref. \cite{johannes}.
Therefore, the study of  quasi 1d quantum HAFM's, i.e. systems  where the in-chain couplings are larger than 
the inter-chain couplings,
are - on one hand - of basic interest in connection with the dimensional 
crossover from one dimensions to two, and on the other hand, 
of interest for the interpretation of  experiments. The  quasi 1d spin-$1/2$ HAFM  has been studied in several papers 
\cite{akai89,parola,ihle,neto,sandvik,affleck,wang,kim,matsumoto01} in recent years. A main focus of these 
studies has been on the estimation of the 
critical inter-chain 
coupling $J_{\perp}^c$ where the transition between the phase with magnetic LRO and
the magnetically disordered phase takes place. 
The answers given in the literature  to this question 
are contradictory and not completely conclusive. 
While some papers find indications for  a 
finite $J_{\perp}^c$ \cite{parola,ihle,neto}, 
others find $J_{\perp}^c=0$
\cite{akai89,sandvik,affleck,wang,kim,matsumoto01} which seems to be more plausible, 
given that the 
GS  of the 1d spin-$1/2$ HAFM
is not gapped. {In particular, data obtained by the quantum Monte Carlo
method (QMC)~\cite{sandvik,kim,matsumoto01}, which is precise for unfrustrated spin models, strongly
support the result $J_{\perp}^c=0$.}
 The behaviour of the quasi 1d spin-$1$ HAFM might be different from its 
spin-$1/2$ counterpart, but is less
studied so far. Indeed, the existing studies of the spin-$1$ case find 
indications for a finite $J_{\perp}^c$
\cite{akai89,kim,koga00,matsumoto01,alet02,wang03} which 
can be attributed to the spin gap between the singlet 
GS and the magnetic excitations.
In Ref.~\cite{akai89} a lower bound of $J_{\perp}^c$ was estimated as 
$J_{\perp}^c \geq 0.025 J_1$, whereas an upper bound  $J_{\perp}^c
\le 0.18 92 J_1$ was given in Ref.~\cite{wang03}. 
{Recent QMC calculations~\cite{kim,matsumoto01,alet02}  yield 
$J_{\perp}^c  \approx 0.043 - 0.044 J_1$, which is only about 10\% of the Haldane gap
$\Delta$.}

To get experimental input for the theoretical work, materials with a quasi 1d behavior 
and a small coupling ratio between inter- and in-chain exchange are needed. 
\\Experimentally, materials such as $Sr_2CuO_3$, $Ca_2CuO_3$~\cite{kojima,rosner}, $Sr_2V_3O_9$ ~\cite{kaul},  $BaCu_2Si_2O_7$~\cite{kenzelmann}, $Sr_2Cu(PO_4)_2$, 
$Ba_2Cu(PO_4)_2$~\cite{belik}  are  quite  good examples of quasi 1d spin-$1/2$ 
antiferromagnets with "not-too-large"
inter-chain coupling.
The spin-$1/2$
antiferromagnet showing the most perfect 1d 
behavior so far is $SrCu(PO_4)_2$ \cite{johannes}; this material has the smallest  ratio 
$k_BT_N/J_1 \sim 6 \times 10^{-4}$ between the N\'eel temperature $T_N$  
and the in-chain coupling $J_1$.
All of the above-mentioned  materials show magnetic LRO below $T_N$, 
although their coupling ratio is remarkably small. Contrary to the spin-$1/2$ materials in  
some quasi-1d spin-$1$ antiferromagnets 
such as  $Ni(C_2H_8N_2)_2NO_2(ClO_4)$ \cite{rennard,aydmer} and other 
$Ni$-compounds, see  Ref.~\cite{rennard},  no 
N\'eel  LRO has been observed as measured down to very low temperatures.

Another interesting system is the 1d spin-$1/2$ HAFM with frustrated second neighbor 
interaction $J_2>0$. 
At $J_2= 0.2411 J_1$ \cite{nomura} a transition to a dimerized state with a spin  gap and
an exponential decay of the spin-spin correlation occurs. 
Hence, for  $J_2>0.2411J_1$ the effect of the inter-chain coupling 
on the GS behavior may be different from the case with $J_2<0.2411J_1$. 
This problem has  not been discussed in detail in the literature so far. 

In the present paper we apply  the  coupled cluster  me\-thod  (CCM) \cite{coester,bishop91} to study the GS LRO of quasi 
1d HAFM with spin quantum numbers $s=1/2$ and $s=1$. 
This approach
is a universal and powerful method of quantum many-body theory. 
{Though  the CCM  is a fairly new method in the field of quantum
spin systems, in recent years it has been
developed to higher levels of
approximation which allows its application to   quantum spin systems with
more and more 
success (for  recent reviews, see Refs.~\cite{farnell04,richter06}).
In particular, the CCM  has the
advantage that it can be applied to frustrated quantum spin systems with
arbitrary dimensions.  
Though, concerning the precision of the results,  
the CCM at the present level of approximation  probably cannot compete with 
the QMC, it allows to find new
results for frustrated systems, for which the QMC fails due to the 
so-called sign problem. 
On the other hand, the comparison of CCM results with 
QMC data for the unfrustrated
models can be considered as a benchmark test of the CCM and is therefore of interest from 
an applied  method point of view. }

\begin{figure}
\begin{center}
\scalebox{0.35}{
\includegraphics{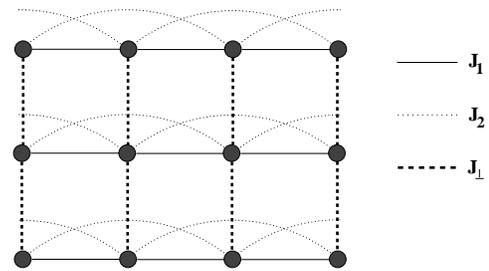}
}
\caption{\label{figure1} Illustration of the quasi one-dimensional HAFM with
the in-chain nearest-neighbor bonds $J_1$, the
frustrating in-chain next-nearest neighbor bonds $J_2$, and the inter-chain
bonds $J_\perp$, cf. the Hamiltonian (\ref{eq1.1}).
All bonds
are antiferromagnetic.}
\end{center}
\end{figure}

The Hamiltonian of the quasi 1d frustrated HAFM  
is written as 
\bea\label{eq1.1}H&=& \sum_n 
\sum_i \big (
J_1{\bf s}_{i,n}\cdot{\bf s}_{i+1,n} + J_2 {\bf s}_{i,n}\cdot{\bf
s}_{i+2,n}\big) \nonumber\\
&+& \sum_i\sum_n J_{\perp} {\bf s}_{i,n}\cdot{\bf s}_{i,n+1}.
\eea
The index $n$ labels the chains and  $i$ the lattice sites within a chain $n$. 
The model is illustrated in Fig.~\ref{figure1}. While for spin quantum
number $s=1$ we consider only the model without frustration ($J_2=0$), 
for $s=1/2$, we will discuss both cases,  $J_2=0$ and $J_2>0$.

\section{The coupled Cluster Method (CCM)}
\label{ccm}
In this section, the CCM formalism will first briefly be outlined. 
For further details the interested reader is referred to 
Refs.~\cite{farnell04,richter06,zeng98,bishop00,krueger00,farnell02,farnell05,rachid05,rachid06,schmal06,bishop07}.
The starting point for the CCM calculation is the choice of a normalized reference or model state
$|\Phi\rangle$, together with a set of (mutually commuting) multi-configurational creation 
operators $\{ C_L^+ \}$ and 
the corresponding set of their Hermitian adjoints $\{ C_L \}$,
\begin{eqnarray}
\label{eq2.1} \langle \Phi|C_L^+ = 0 = C_L|\Phi \rangle \quad \forall L\neq 0, \quad C_0^+\equiv 1 \\
\label{eq2.2}[C_L^+,C_J^+] = 0 =[C_L,C_J] \; .
\end{eqnarray}
The operators $C_L^+$ ($C_L$) are defined over a complete set of many-body configurations 
denoted by the set-indices $\{ L \}$. 
For the set $\{|\Phi\rangle, C_L^+\}$ the CCM parametrization of the exact ket and bra GS eigenvectors
$|\Psi\ra$ and $\la\tilde{\Psi}|$ of our many-body system are given  by
\begin{eqnarray}\label{eq5} 
|\Psi\ra=e^S|\Phi\ra \; , \mbox{ } S=\sum_{L\neq 0}a_LC_L^+ \;\\
\label{eq5b}
\la \tilde{ \Psi}|=\la \Phi |\tilde{S}e^{-S} \; , \mbox{ } \tilde{S}=1+ \sum_{L\neq 0}\tilde{a}_LC_L \; .
\end{eqnarray}
The CCM correlation operators, $S$ and $\tilde{S}$, contain the correlation coefficients, $a_L$ and $\tilde{a}_L$, 
which have to be calculated.  Once these values are known, all the GS properties of the many-body system can be derived from them. To find the GS correlation coefficients $a_L$ and $\tilde{a}_L$,
we simply require that the expectation value $\bar H=\la\tilde\Psi|H|\Psi\ra$
(GS energy) is a minimum with respect
to the entire set $\{ a_L , \tilde{a}_L \}$, which leads to the GS CCM ket-state and bra-state
equations
\begin{eqnarray}
\label{eq6}
\langle\Phi|C_L^-e^{-S}He^S|\Phi\rangle = 0 \;\; ; \; \forall L\neq 0\\ 
\langle\Phi|{\tilde S}e^{-S}[H, C_L^+]e^S|\Phi\rangle = 0 \; \; ; \; \forall L\neq 0.
\end{eqnarray}

For the spin systems considered herein, we choose 
the N\'eel state with spins aligned in 
the $z$-direction as the reference state. 
The reasoning behind this choice is evident for the unfrustrated case,
[~\cite{bishop00}], since the N\'eel state is the classical GS for
$J_2=0$. 
For the $s=1/2$ case, the frustrated model (i.e. $J_2 > 0$) is considered
below as well.
Note that the classical GS is an incommensurate spiral state for $J_2 >
0.25J_1$.
However, 
the quantum GS for $J_\perp=0$ does   not exhibit spiral ordering for values
of $J_2$ less than
$J_2 \sim 0.5 J_1$, 
but it is rather a collinear state.~\cite{richter06,bursill,white,aligia}
Hence, the N\'eel state is an appropriate reference state for $0 < J_2
\lesssim 0.5 J_1$ as well.~\cite{richter06}

To treat each side equivalently,
we perform a rotation of the local axis of the up spins, such 
that all spins in the reference state
align in the negative $z$-direction. 
In this new set of local spin coordinates 
the reference state and the corresponding creation operators $C_L^+$ are given by
\begin{equation}
\label{set1} |{\hat \Phi}\ra = |\downarrow\downarrow\downarrow\downarrow\cdots\rangle \; ; \mbox{ } C_L^+ 
= {\hat s}_i^+ \, , \, {\hat s}_i^+{\hat s}_j^+ \, , \, {\hat s}_i^+{\hat s}_j^+{\hat s}_k^+ \, ,\, \ldots \; ,
\end{equation}
where the indices $i,j,k,...$ denote arbitrary lattice sites.
For the discussion of the GS N\'eel LRO,  we have to calculate 
the order parameter (sublattice magnetization) $M$, which is  
given within the CCM scheme by 
\begin{equation}
\label{eqnm}M= -\frac{1}{N}\la\tilde{\Psi}|\sum_{i=1}^N{\hat s}_i^z|\Psi\ra.
\end{equation}

The CCM formalism becomes exact if we take into account all possible multispin 
configurations for 
the correlation operators $S$ and $\tilde S$.  However, in general, this is  
impossible to do in practice for  a many-body quantum system. 
It is therefore necessary to use approximation 
schemes in order to truncate the expansions of $S$ and $\tilde S$ in 
Eqs.~(\ref{eq5},\ref{eq5b}) in any practical calculation.  
A very general  approximation scheme is the so-called SUB$n$-$m$ approximation. In this approximation, all correlations 
in the correlation operators $S$ and $\tilde S$ are taken into account, as long as they span a range of no more than 
$m$ contiguous sites and contain only  $n$ or fewer spins. 
In most cases, however, the SUB$n$-$n$ scheme is used (i.e., with $n=m$), and in these cases, it is referred to as the
LSUB$n$ scheme (for spin-$1/2$ systems). 
To find all the different (i.e.  fundamental) 
configurations entering $S$ and $\tilde S$ for a given level of
SUB$n$-$n$ approximation, we use the lattice symmetries.

\begin{figure}
\scalebox{0.65}{\includegraphics{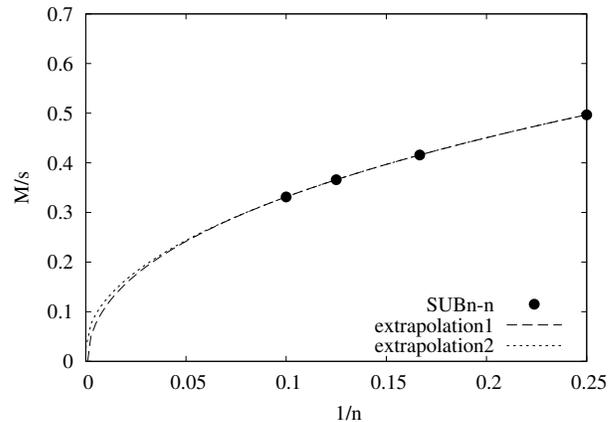}}
\caption{\label{figure2}{ Extrapolation of the
GS order parameter (sublattice magnetization) $M$
scaled by the spin quantum number $s$ for the pure unfrustrated $s=1/2$ chain
(i.e.
$J_\perp=0$ and $J_2=0$). The two methods of extrapolation corresponding to Eqs.
(\ref{scal_m1}) and (\ref{scal_m2}) are indicated by
'extrapolation1' and 'extrapolation2', respectively. } }
\end{figure}

Although there is no  theory available for how the results of 
the SUB$n$-$n$ approximations scale with $n$, there is nevertheless 
a great deal of experience in how to extrapolate the raw CCM SUB$n$-$n$ data properly 
to $n \to
\infty$.~\cite{richter06,zeng98,bishop00,krueger00,rachid05,schmal06,bishop07}
The best results for the extrapolation of the order parameter are obtained if the poor 
SUB2-2 data are omitted.
Previously, for systems showing an order-disorder quantum phase transition
~\cite{richter06,rachid05,schmal06}, 
 a leading 'power-law' extrapolation for the order parameter 
\be \label{scal_m1}
 M(n)=a_0+a_1\left(\frac{1}{n}\right)^{a_2} \, ,
\ee
has been used successfully to determine the phase transition points.
In Eq. (\ref{scal_m1}) 
the leading exponent $a_2$ is determined directly from
the SUB$n$-$n$ data. Alternatively, as has been discussed recently in
Ref.~\cite{bishop07}, 
an extrapolation scheme with a fixed exponent but including an
additional power in $1/n$, i.e.
\be \label{scal_m2}
 M(n)=b_0+\left(\frac{1}{n}\right)^{1/2}\left(b_1+b_2\frac{1}{n}\right ) \, ,
\ee
can also be used to find the phase transition point.
The extrapolation of the order parameter is illustrated in Fig.~\ref{figure2}
for one particular data set - namely, for the unfrustrated
$s=1/2$ chain with zero inter-chain coupling, i.e. at the expected critical
point. 
Note that when using
the extrapolation of Eq.~(\ref{scal_m1}), the exponent $a_2$ for the considered data
set is $a_2 =
0.414 $, which is a value not far from the 
 fixed leading exponent $1/2$ used in Eq.~(\ref{scal_m2}).      
Below, we will use both extrapolation formulas to determine the critical
inter-chain coupling 
$J_{\perp}^c$, which will naturally yield
slightly different values for $J_{\perp}^c$.
We will use this difference as an estimation of the reliability of our results.

Using parallel processing \cite{farnell05,ccm} 
we are able to use the CCM up to the SUB$10$-$10$ approximation 
for the spin-$1/2$ system, and up to the SUB$8$-$8$
approximation for the spin-$1$ system, which corresponds  
to the solution of more than $10^4$ coupled nonlinear equations.
\begin{figure}
\scalebox{0.65}{\includegraphics{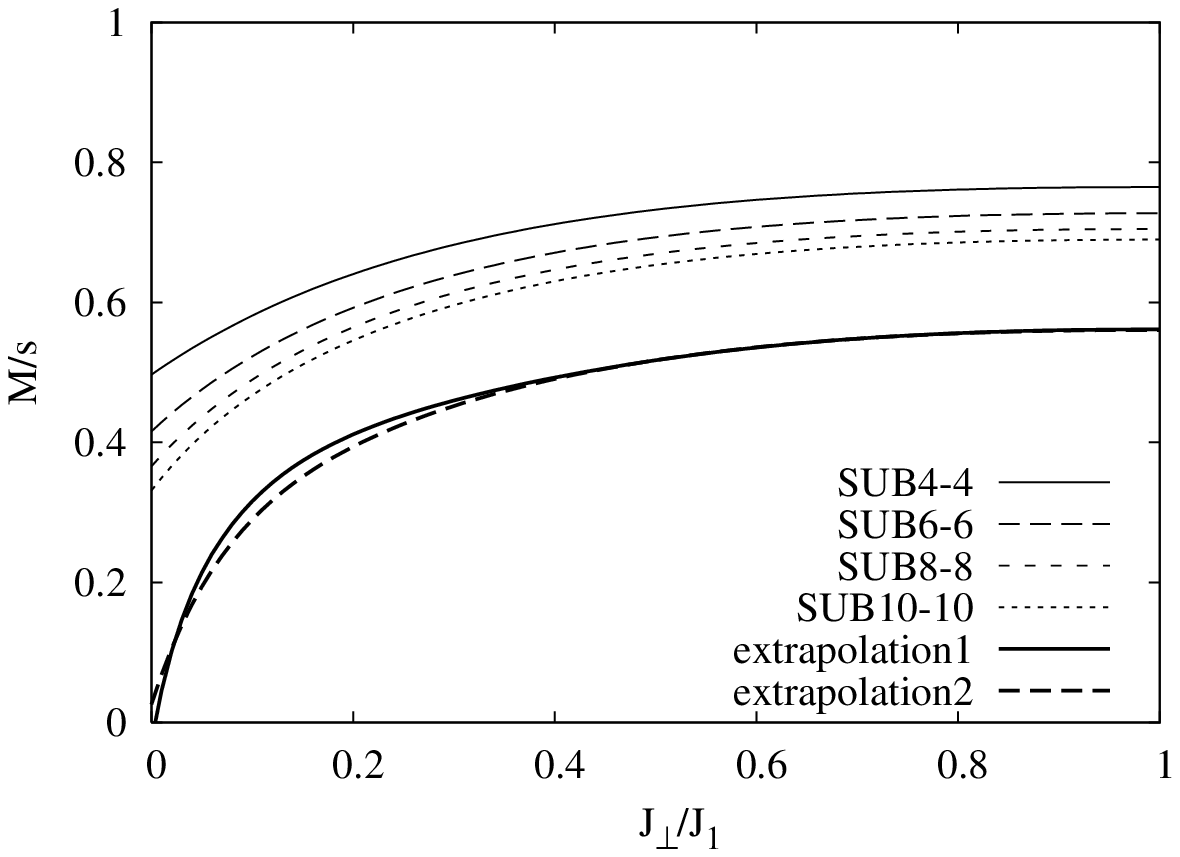}}
\caption{\label{figure3}{  The dependence of the GS order parameter (sublattice magnetization) $M$
scaled by the spin quantum number $s$ on the inter-chain coupling 
$J_\perp$ 
for the unfrustrated $s=1/2$ quasi 1d HAFM.  The two methods of extrapolation corresponding to Eqs.
(\ref{scal_m1}) and (\ref{scal_m2}) are indicated by
'extrapolation1' and 'extrapolation2', respectively.}}
\end{figure}
\begin{figure}
\scalebox{0.65}{\includegraphics{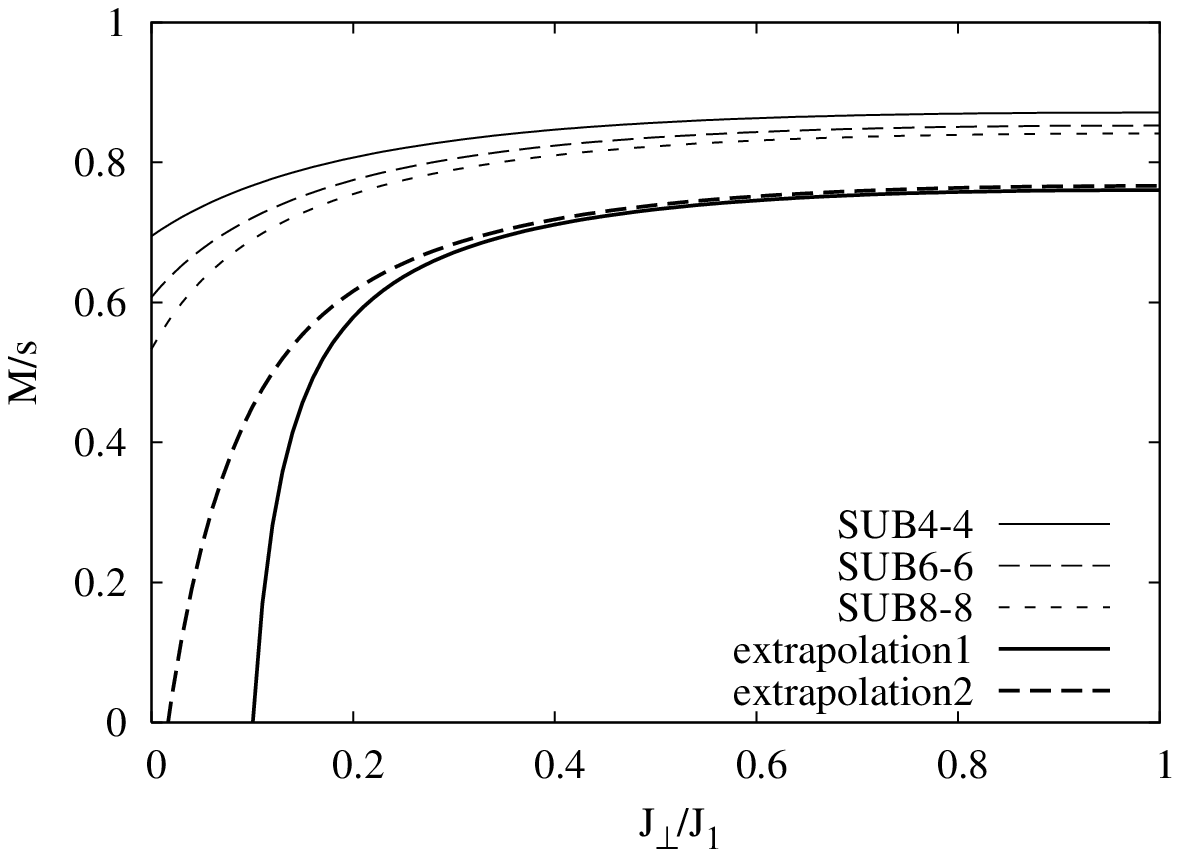}}
\caption{\label{figure4} { 
The dependence of the GS order parameter (sublattice magnetization) $M$
scaled by the spin quantum number $s$ on the inter-chain coupling 
$J_\perp$ 
for the unfrustrated $s=1$ quasi 1d HAFM.  The two methods of extrapolation corresponding to Eqs.
(\ref{scal_m1}) and (\ref{scal_m2}) are indicated by
'extrapolation1' and 'extrapolation2', respectively.} }
\end{figure}
\begin{figure}
\scalebox{0.65}{\includegraphics{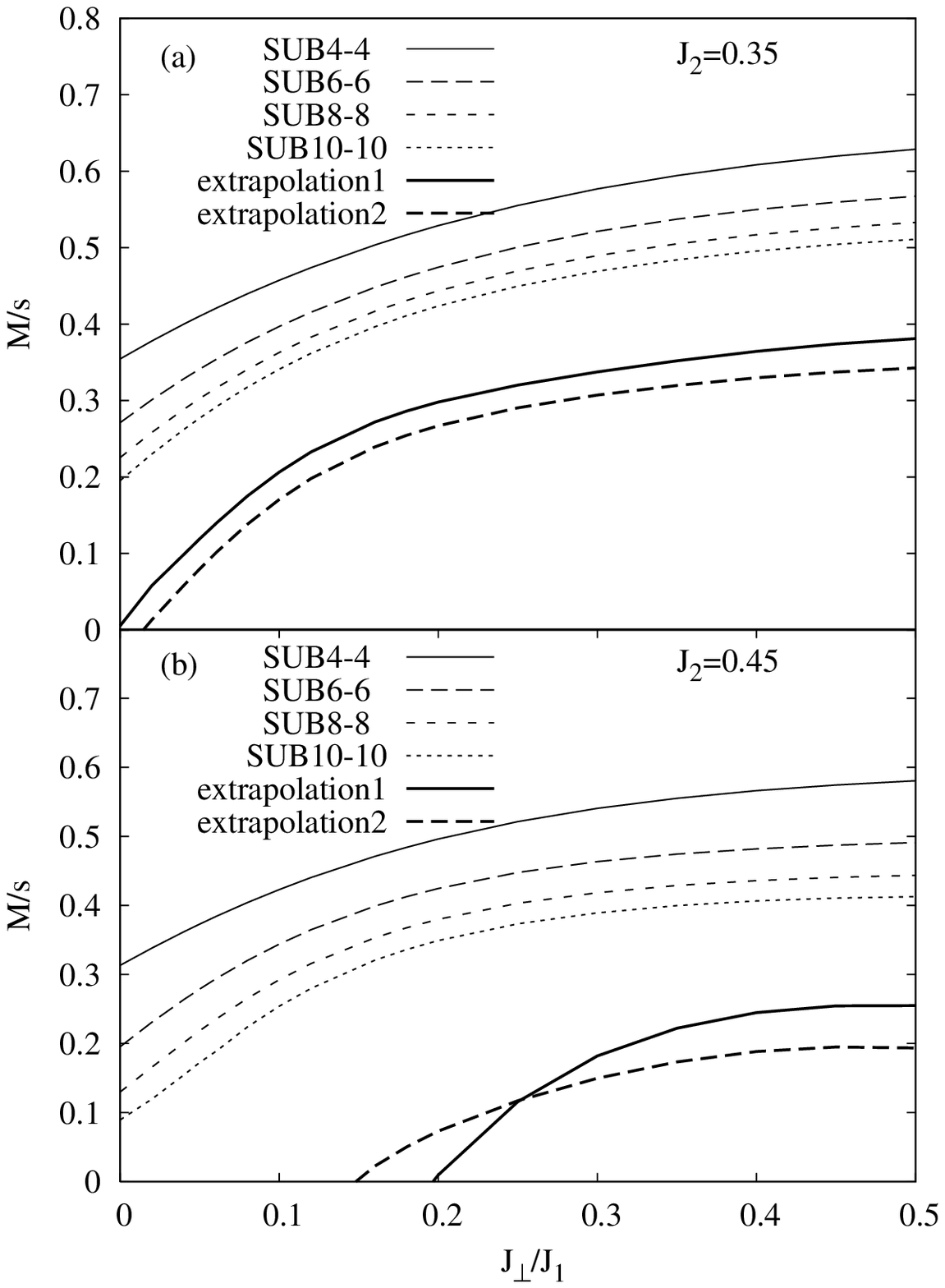}}
\caption{\label{figure5}{  The GS order parameter (sublattice magnetization) $M$ scaled by the spin 
quantum number $s$ in dependence on the inter-chain coupling $J_\perp$ for the frustrated 
$s=1/2$ quasi 1d HAFM.  The two variants of extrapolation according to Eqs.
(\ref{scal_m1}) and (\ref{scal_m2}) are indicated by
'extrapolation1' and 'extrapolation2', respectively.
a: $J_2=0.35J_1$; b: $J_2=0.45J_1$ .}}
\end{figure}
\section{Results}
The results for the dependence of the order parameter $M$ scaled by the spin quantum number $s$
on the inter-chain coupling $J_\perp$  are shown in
Figs.~\ref{figure3}--\ref{figure5}. In those figures, we show the raw CCM
SUB$n$-$n$ data used
for the extrapolation, as well as the results for the extrapolation $n \to
\infty$, cf. Sect.~\ref{ccm}.

For the unfrustrated quasi 1d HAFM with $s=1/2$ we find 
that for $J_1 \ge J_\perp \gtrsim 0.2J_1$,  
the variation of the order parameter $M$ with $J_\perp$ is small, see
Fig.~\ref{figure3}.
Only for $J_\perp < 0.2J_1$ is the sublattice magnetization significantly
diminished. {As demonstrated in Fig.~\ref{figure3}, it is evident that both extrapolation
schemes
lead to similar results.
Using the extrapolation formula Eq.~(\ref{scal_m1}),
we obtain 
the critical inter-chain coupling $J_\perp^c \sim 0.003J_1$ at which the N\'eel LRO
disappears . On the other hand, the extrapolation based on
Eq. (\ref{scal_m2}) leads to a finite but very small order parameter $M \approx
0.01$.  These results are consistent with the conclusion      
that for the unfrustrated spin-$1/2$ system, 
an infinitesimally small inter-chain coupling is sufficient to stabilize 
antiferromagnetic LRO.} This statement  supports 
the findings of Refs. \cite{akai89,sandvik,affleck,wang,kim} and are related 
to the gapless GS of the strictly 1d $s=1/2$ HAFM. 

Next, we consider the unfrustrated $s=1$ HAFM. Here the number of 
fundamental configurations in the CCM SUB$n$-$n$ approximation is much larger
than for $s=1/2$, and we 
are able to present CCM data  up to $n=8$. {Hence, 
the extrapolation for $s=1$ is expected to be less reliable than for $s=1/2$, for which
CCM data
up to $n=10$ are available.}
Again, 
for $J_1 \ge J_\perp \gtrsim 0.2 J_1$, 
the variation of $M$ with $J_\perp$ is small, see Fig.~\ref{figure4}. 
However, $M/s$ is significantly
larger than as calculated for $s=1/2$, 
indicating the decrease of quantum fluctuations
in the square-lattice HAFM with increasing spin quantum number $s$. 
A strong reduction of the order parameter occurs for $J_\perp \sim
0.2J_1$. {Finally, $M$ vanishes at a critical value  
$J_\perp^c > 0$, i.e. a finite inter-chain coupling is necessary 
to stabilize GS N\'eel LRO. However, both extrapolation schemes lead to
different numerical values for $J_\perp^c$, namely $J_\perp^c \sim 0.1J_1 $ 
using Eq.~(\ref{scal_m1}) but $J_\perp^c \sim 0.02J_2 $ 
using Eq.~(\ref{scal_m2}). Both values differ by approximately a factor of 2 from
the QMC results~\cite{kim,matsumoto01,alet02}
$J_{\perp}^c  \approx 0.043 - 0.044 J_1$. This difference gives an
indication of the accuracy of the CCM including SUB$n$-$n$ data up to $n=8$. }

Next, we consider the frustrated
quasi 1d
 $s=1/2$ HAFM  {for which QMC calculations are not possible.} 
The classical GS for $J_2>0.25J_1$ is  an incommensurate spiral state independent of the value of
$J_\perp$.
In contrast to the classical model, for the strictly 1d problem ($J_\perp=0$) the quantum  GS 
 does not exhibit spiral correlations for $0.25J_1 < J_2 \lesssim 0.5J_1$ 
\cite{richter06,bursill,white,aligia}, but is 
gapped with a spin gap $\Delta$ strongly varying with $J_2$.~\cite{white,chitra}
However, in presence of an appreciable inter-chain coupling $J_\perp$, 
spiral
correlations may appear in the quantum model as well \cite{richter06,zinke}.
Therefore, we restrict our calculations
to $J_\perp < 0.5J_1$ and to $J_2 \le 0.46J_1$,
where the CCM works well when based on a collinear reference state.
The spin gap $\Delta$ of the 1d problem ($J_\perp=0$) was found to be 
very small for $J_2 < 0.4 J_1$, but $\Delta$ increases rapidly between $0.4J_1 <
J_2 <0.5J_2$.~\cite{white,chitra}
\begin{figure}
\scalebox{0.65}{\includegraphics{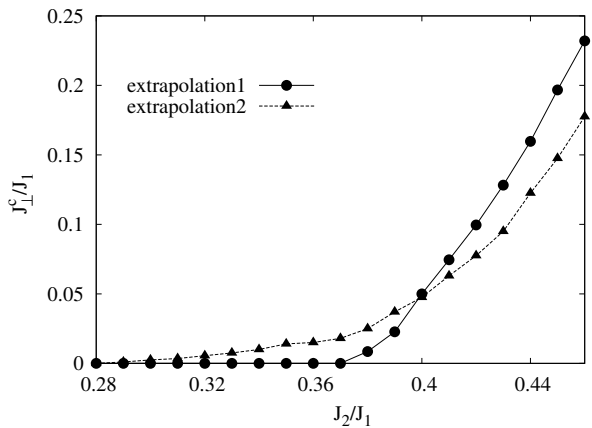}}
\caption{\label{figure6} { The critical inter-layer coupling $J_\perp^c$
in dependence on the frustration parameter  $J_2$ for the
frustrated $s=1/2$ quasi 1d HAFM.  The two variants of extrapolation according to Eqs.
(\ref{scal_m1}) and (\ref{scal_m2}) are indicated by
'extrapolation1' and 'extrapolation2', respectively.}}
\end{figure}
\begin{figure}
\scalebox{0.65}{\includegraphics{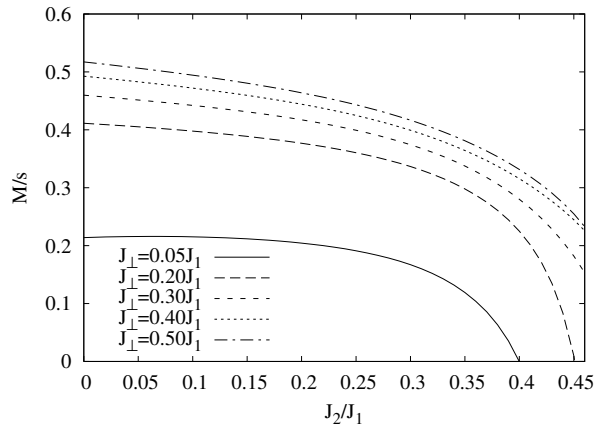}}
\caption{\label{figure7} The GS order parameter (sublattice magnetization)
$M$  scaled by the spin quantum 
number $s$  for the frustrated $s=1/2$ quasi 1d HAFM in dependence on the frustration parameter $J_2$
and for various strengths of the inter-chain coupling $J_\perp$. }
\end{figure}

The results for $M/s$ versus $J_\perp$ for $J_2=0.35J_1$ and $J_2=0.45J_2$
are shown in Fig.~\ref{figure5}. { For  
$J_2=0.35J_1$ the critical inter-chain coupling is $J^c_\perp \sim 0$ when using the extrapolation Eq.~(\ref{scal_m1})
and is $J^c_\perp \sim 0.014J_1$ when using the extrapolation Eq.~(\ref{scal_m2}). 
Knowing that for $J_2=0.35J_1$ and $J_\perp=0$, the spin gap should be finite~\cite{nomura}
but very small~\cite{white,chitra},  a zero or small $J^c_\perp$ is reasonable.
On the other hand, for    
 $J_2=0.45J_1$, where the spin gap for $J_\perp=0$ is $\Delta \sim
0.13J_1$~\cite{white,chitra} we already obtain quite a large value of  $J^c_\perp \sim
0.20 J_1$ using the 
extrapolation of Eq.~(\ref{scal_m1}) and $J^c_\perp \sim 0.15 J_1$ using the extrapolation
of Eq.~(\ref{scal_m2})).
Interestingly, the ratio $J_\perp^c/\Delta$ is seems to be larger than for the
unfrustrated $s=1$ HAFM. 
The variation  of $J_\perp^c$ with $J_2$ 
is shown in Fig.~\ref{figure6}. Both extrapolation schemes yield 
qualitatively similar results. The variation of $J^c_\perp$ with $J_2$ 
is quite similar to the variation of the spin gap with
$J_2$~\cite{white,chitra}.} However, we observe a monotonic increase of the
ratio
$R=J_\perp^c/\Delta$ from $R \sim 0.8$ to $R \sim 1.6 $ in the region
$0.4J_1 \le J_2 \le 0.46J_2$.

The variation of the order parameter  $M/s$ with frustration is illustrated in
Fig.~\ref{figure7}, {where for the sake of clarity, only the data 
obtained using an extrapolation scheme
Eq.~(\ref{scal_m1}) are shown.}  
As expected, frustration weakens the magnetic order and $M$ becomes smaller
with increasing $J_2$. Obviously, for fixed but not too large inter-chain coupling
$J_\perp$, a quantum phase transition between N\'eel
LRO and a magnetically disordered phase can be driven by frustration. However,
In similarity to recent findings for the quasi-2d $J_1-J_2$
model~\cite{schmal06}, it is likely that for stronger inter-chain coupling 
no magnetically disordered phase appears.

Let us finally mention that the quantum phase transition
in the frustrated model is interesting from a more general point of view. 
For $0.2411J_1 < J_2 \lesssim 0.5J_1$, the model exhibits
two ground state
phases, each breaking different symmetries, namely (i) the rotationally invariant,
spontaneously dimerized 
phase for zero
(or small) $J_\perp$ breaking the translational symmetry of the lattice, and
(ii) the N\'eel phase, breaking the spin rotational symmetry.
A continuous transition
between the dimerized phase and the N\'eel
ordered phase 
is prohibited within the Landau theory.~\cite{senthil04}. Hence three
different scenarios are possible. First, that there is a (small) disordered featureless 
spin-liquid 
phase between the dimerized and the N\'eel
phase;
second, that there is a first order transition between the dimerized and the N\'eel
phase, and, third and most interesting, that the above-described transition is a candidate
for a deconfined quantum critical
point.~\cite{senthil04} This question cannot be answered within the current approach
but deserves further consideration.

To summarize, we find that the transition from the non-magnetic 
1d GS to the magnetically ordered 2d GS 
in quasi 1d quantum HAFM's  can be well described by the CCM if higher orders of
approximation are used.
Our results indicate that this transition, driven by the inter-chain coupling
$J_\perp$,
is related to the excitation gap of the strictly 1d HAFM, i.e. at
$J_\perp=0$. If the 1d quantum GS
is gapless, most likely the magnetic LRO sets in immediately when $J_\perp$ is
switched on, whereas for gapped GS's, a finite $J_\perp$ is necessary to
establish magnetic LRO in the GS.

{{\it   Acknowledgment:}
This work was supported by the DFG (project Ri615/16-1).}


\end{document}